\documentclass{JHEP3}
\usepackage{graphicx}
\usepackage{amssymb}
\usepackage[mathscr]{eucal}
\usepackage{amsmath}
\usepackage{cite}
\usepackage{afterpage}

\newcommand{\gsimm}{\raise.3ex\hbox{$>$\kern-.75em\lower1ex\hbox{$\sim$}}}
\newcommand{\lsimm}{\raise.3ex\hbox{$<$\kern-.75em\lower1ex\hbox{$\sim$}}}

\newcommand{\be}{\begin{equation}}
\newcommand{\ee}{\end{equation}}
\newcommand{\ba}{\begin{eqnarray}}
\newcommand{\ea}{\end{eqnarray}}

\newcommand{\bea}{\begin{eqnarray*}}
\newcommand{\eea}{\end{eqnarray*}}

\title{Cosmological Tests of the Disformal Coupling to Radiation}
\author{Philippe Brax \\
  Institut de Physique Th\'eorique, CEA, IPhT, CNRS, URA 2306,
  F-91191Gif/Yvette Cedex, France \\ E-mail:
  \email{philippe.brax@cea.fr}}
 \author{Clare Burrage\\
 School of Physics and Astronomy, University of Nottingham, Nottingham, NG7 2RD, United Kingdom
  \\ E-mail:
  \email{Clare.Burrage@nottingham.ac.uk} }

\author{Anne-Christine Davis \\ Department of Applied Mathematics and Theoretical Physics,
Centre for Mathematical Sciences, Cambridge CB3 0WA, UK\\ E-mail:
\email{a.c.davis@damtp.cam.ac.uk}}
\author{Giulia Gubitosi\\ Dipartimento di Fisica, Universit\`a di Roma ``La Sapienza'' and INFN sez. Roma1, P.le Aldo Moro 2, Roma, Italy
   \\ E-mail:
  \email{giulia.gubitosi@roma1.infn.it}}
\date{today}
\abstract{Light scalar fields can naturally couple disformally to Standard Model fields without giving rise to the unacceptably large fifth forces usually associated with light scalars.  We show that these scalar fields can still be studied and constrained through their interaction with photons, and focus particularly on changes to the Cosmic Microwave Background spectral distortions and violations of the distance duality relation.  We then  specialise our constraints to scalars which could play the role of  axionic quintessence.}

\begin{document}
\section{Introduction}
New, light scalar degrees of freedom seem to be an almost inevitable consequence of attempts to explain the current acceleration of the expansion of the Universe.  Whether they are included in the theory explicitly as quintessence, or k-essence fields (for a review see \cite{Copeland:2006wr}), or whether they are a consequence of attempts to modify gravity such as in $f(R)$  and massive gravity  theories (for a review of modified gravity models of interest to cosmology see \cite{Clifton:2011jh}) their presence typically causes new headaches for  model building theorists.

Firstly, if a new light scalar field couples to matter fields it is expected that it will mediate a new, long range fifth force.  Such fifth forces are excluded by terrestrial and solar system measurements, unless their couplings to matter are extremely weak compared to gravity; a new fine tuning problem.  In order to be in agreement with these measurements we either need a reason why such couplings between the scalar field and matter are forbidden, or we need to make the theory non-linear in such a way that it has a screening mechanism which dynamically suppresses the effects of the scalar force.

Secondly, if these scalar fields are related to the mechanism driving the accelerating expansion of the Universe, we expect them to have Compton wavelengths corresponding to distance scales similar to the size of the observable Universe today.  The introduction of such a small mass scale for the scalar field is the source of a new fine tuning problem.  If the scalar couples to matter fields, we would typically expect  matter loop corrections to renormalise this mass to higher energy scales, and therefore if we are not prepared to accept the necessary fine tuning to keep the mass light we must invoke a new mechanism to explain it.

There is one simple thing that can be done  to address both of these problems and that is to insist that the action for the scalar field is invariant under the  shift symmetry $\phi\rightarrow \phi + c$, for a constant $c$.  This symmetry can be either exactly respected or softly broken; when the symmetry is respected both a mass term for the scalar field and the interactions between the scalar field and matter that give rise to fifth forces are forbidden.  If the symmetry is softly broken then we have a natural explanation for why the mass terms and the couplings to matter are small. One prime example is provided by the Goldstone modes of a global symmetry where the interaction potential results from a soft and explicit breaking of the symmetry. In the appendix we will present models of axion quintessence which fall in this category and describe thawing models where the late time acceleration of the expansion of the Universe comes from the dynamics of the scalar field moving very little from its initial value in the early Universe.

Imposing a shift symmetry does not, however, forbid all interactions between the scalar field and matter.  In particular it allows for the so-called disformal couplings between matter and a scalar field.  These interactions were first discussed by Bekenstein \cite{Bekenstein:1992pj}, who showed that the most general metric that can be constructed from $g_{\mu\nu}$ and a scalar field that respects causality and the weak equivalence principle is;
\begin{equation}
\tilde{g}_{\mu\nu}=A(\phi,X)g_{\mu\nu} + B(\phi,X) \partial_\mu \phi \partial_\nu \phi\; ,
\label{eq:bekmetric}
\end{equation}
where the first term gives rise to `conformal' couplings between the scalar field and matter, and the second term is the `disformal' coupling.  Here  $X=(1/2)g^{\mu\nu}\partial_{\mu}\phi\partial_{\nu}\phi$.
As we will see the disformal interactions  give rise to Lagrangian interaction terms of the form
\begin{equation}
\mathcal{L} \supset \frac{1}{M^4}\partial_\mu \phi\partial_\nu \phi T^{\mu\nu}\;.
\end{equation}
where $T^{\mu\nu}$ is the energy momentum tensor of matter fields.  It is thus easy to see why the scalar field can couple to matter in this way and not give rise to a fifth force.  The interactions controlled by this coupling involve two copies of the scalar field, and at least two matter particles.  Therefore the first scalar corrections to two particle scattering must be at the one-loop level and so this type of coupling for the scalar field does not give rise to any classical forces in vacuum. This is all the more true that at the classical level, in static configurations, the coupling to the matter energy density vanishes annulling any fifth force which only exists in dynamical situations.

Just because these couplings do not give rise to fifth forces does not mean that their interactions with Standard Model fields are undetectable.  The disformal interactions of scalars with photons  can be probed in laboratory experiments \cite{Brax:2012ie}. In
models motivated by Galileon theories and massive gravity  constraints have been
put on the disformal interactions from studying gravitational lensing and the velocity dispersion of
galaxies \cite{Wyman:2011mp,Sjors:2011iv}.  Other cosmological implications of disformal scalars have been considered in \cite{Zumalacarregui:2010wj,Koivisto:2012za}. In this work we focus on what cosmological observations can tell us about the possible existence of disformal scalar fields.  This connects with the work in \cite{vandeBruck:2013yxa}, however in that work only the interactions between scalar fields and radiation were studied and all effects of matter were neglected.

It turns out that disformally coupled models share a strong link with varying speed of light theories\cite{Magueijo:2003gj}\footnote{In the early universe, disformally coupled models are also known to solve horizon and flatness problems and produce scale-invariant density fluctuations, without the need of an inflationary mechanism \cite{Clayton:2001rt,Magueijo:2008sx}.}. Indeed we find that light rays in these models follow geodesics where the speed of light is not constant anymore. This implies that distance measurements are affected and more particularly the distance duality relation. In this work we show how the reciprocity relation is altered. When adding a direct coupling to electromagnetism and therefore a change in time of the fine structure constant, we find that the duality relation receives corrections of two origins. One follows from the attenuation of light along geodesics and is associated to the variation of the fine structure constant, another one is due to the variation of the speed of light and disturbs geodesics. We also find that the amplitude of the CMB spectrum is  modified by the same quantity as the duality relation. This can be constrained as it leads to a $\mu$ distorsion of the CMB spectrum which is precisely bounded since last scattering. Together with quasar constraints on the duality relation, we can therefore give new bounds on a combination of the variation of the speed of light and the fine structure constant from last scattering and from a redshift $z\sim 1$. When complementary bounds on the fine structure constant are available, this provides independent constraints on the variation of the speed of light, typically from $z\sim 1$ at the percent level.
Recently, a phenomenological link between a deviation of the CMB temperature and the duality relation was also obtained \cite{Avgoustidis:2013bqa}. Here we show how similar results can be obtained when both a disformal and a direct coupling to photons is included.

The  existence of light scalar fields which interact with matter through a disformal coupling, and possible ways to detect them, is of interest in its own right, and so in this work we will largely remain agnostic about the origin of the scalar. However the study of these scalar fields is particularly interesting in light of their connection to some of the most natural explanations of the current acceleration of the expansion of the Universe, and so in later sections we also relate our constraints to these axionic quintessence models. In this case, we find bounds on $M$ which depend on the equation of state of dark energy now.

In section 2, we discuss the dynamics of disformally coupled scalar fields. In section 3, we study electromagnetism in the presence of a disformal coupling and show that light rays follow geodesics where the speed of light varies. In section 4, we find how the distance duality relation is modified by both a variation
of the speed of light and the fine structure constant. In section 5, the distorsion of the CMB spectrum is calculated and we show that its dependence on the fine structure constant and the speed of light is identical to the deviation of the duality relation from its GR counterpart. Finally in section 6 we give phenomenological constraints on axionic models which are presented in the appendix. We conclude in section 7.

\section{Disformally coupled scalar fields}

We consider the coupling of a scalar field to matter governed by the action
\be
S=\int d^4x \sqrt{-g}\left(\frac{R}{2\kappa_4^2} -\frac{1}{2} (\partial \phi)^2 -V(\phi)\right) + S_m(\psi_i, \tilde g_{\mu\nu})\;,
\label{eq:action}
\ee
where
\be
\tilde g_{\mu\nu}= g_{\mu\nu} +\frac{2}{M^4} \partial_\mu\phi \partial_\nu \phi\;.
\label{eq:tildemetric}
\ee
This is not the most general disformal metric as given by Bekenstein in equation (\ref{eq:bekmetric}), however it describes all the leading order effects of the disformal coupling, and is much simpler to work with.  The coupling scale $M$ is constant and unknown and should be fixed by observations.

The metric $\tilde{g}_{\mu\nu}$ is the Jordan frame metric with respect to which matter is conserved. This follows from diffeomorphism invariance of $S_m$; if
$\delta_\xi \tilde g_{\mu\nu}= \tilde D_\mu \xi_\nu +\tilde D_\nu \xi_\mu$
and $\tilde D_\mu$ is the covariant derivative with respect to $\tilde g_{\mu\nu}$, then
\be
\delta_\xi S_m= -\int d^4x \sqrt{-\tilde g} \xi_\nu \tilde D_\mu \tilde T^{\mu\nu}=0\;,
\ee
 and therefore the corresponding conservation equation (as $\xi_\mu$ is arbitrary) is
 \be
 \tilde D_\mu \tilde T^{\mu\nu}=0\;,
\ee
where the Jordan frame energy momentum tensor is
\be
\tilde T^{\mu\nu}= \frac{2}{\sqrt{-\tilde g}} \frac{\delta S_m}{\delta \tilde g_{\mu\nu}}\;.
\ee
On the other hand, the metric $g_{\mu\nu}$ defines the Einstein frame and in this frame  energy-momentum is not conserved. This can be most easily seen from the Einstein equation
\be
G_{\mu\nu}= \kappa_4^2 (T_{\mu\nu}^\phi + T_{\mu\nu})\;,
\ee
where $G_{\mu\nu}= R_{\mu\nu}-\frac{1}{2} R g_{\mu\nu}$,
\be
T_{\mu\nu}^\phi=\partial_\mu\phi\partial_\nu\phi -g_{\mu\nu}\left(\frac{(\partial\phi)^2}{2} +V(\phi)\right)\;,
\ee
and the Einstein frame energy-momentum for matter is $T^{\mu\nu}= (2/\sqrt{-g})(\delta S_m/\delta g_{\mu\nu})$, so that $\nabla^{\mu}G_{\mu\nu}=0$ implies only that a combination of matter and scalar field is conserved (see the following section).
Other possible choices of frames, and the connection between disformally coupled scalar fields and DBI-Galileon models are discussed in \cite{Zumalacarregui:2012us}.

\subsection{The equations of motion}
The equations of motion for the metric, scalar and matter fields can be derived from the action in equation (\ref{eq:action}).
The Klein-Gordon equation is
\be
\Box\phi -\frac{2}{M^4}D_\mu (\tilde T^{\mu\nu} \partial_\nu \phi) =\frac{\partial V}{\partial \phi}\;.
\label{eq:scalareom}
\ee
The evolution of matter in the Einstein frame is therefore
\be
D^\mu T_{\mu\nu}=-\frac{2}{M^4}\partial^\nu \phi D_\mu(\partial_\lambda \phi T^{\lambda \mu})\;,
\ee
where $D_\mu$ is the covariant derivative with respect to $g_{\mu\nu}$, and we have used the scalar equation of motion (\ref{eq:scalareom}) to simplify this expression. Notice that, as expected, matter is not conserved anymore.

In what follows we will restrict ourselves to the leading order effects of the disformal coupling between the scalar field and matter.  Therefore we calculate only to leading order in $1/M^4$, implying that the Klein-Gordon equation reduces to
\be
\Box\phi -\frac{2}{M^4}D_\mu ( T^{\mu\nu} \partial_\nu \phi) =\frac{\partial V}{\partial \phi}\;,
\ee
and the action can be expanded as
\be
S=\int d^4x \sqrt{-g}\left(\frac{R}{2\kappa_4^2} -\frac{1}{2} (\partial \phi)^2 -V(\phi)+ \frac{1}{M^4} \partial_\mu\phi\partial_\nu\phi T^{\mu\nu}\right) + S_m(\psi_i, g_{\mu\nu})\;.
\ee
The new coupling to matter involves two derivatives and can only be probed in the presence of matter when dynamical situations are considered.
\subsection{Cosmological evolution}
If the Universe is homogeneous and isotropic then it is described in both Einstein and Jordan frames by an FRW metric, with related proper times.
In the Einstein frame we have
\be
ds^2=-dt^2 +a^2 dx^2\;,
\ee
and the associated Jordan frame metric is
\be
ds^2=-\left(1-\frac{2\dot\phi^2}{M^4}\right) dt^2 +a^2 dx^2=-d\tilde t^2+a^2 dx^2\;.
\ee

Cosmologically we consider that the matter in the Universe is well described by a perfect  fluid in the Einstein frame with
\be
T_{\mu\nu}= (\rho+p) u_\mu u_\nu + pg_{\mu\nu}\;,
\ee
where $g^{\mu\nu} u_\mu u_\nu=-1$.
This acts as a source for the scalar field and the Klein-Gordon equation (\ref{eq:scalareom}) becomes
\be
(\ddot \phi + 3H \dot \phi) \left(1-\frac{2p}{M^4}\right) +\frac{2}{M^4} \dot \phi [ \dot \rho+ 3H(\rho+p)]=  -\frac{\partial V}{\partial \phi}\;,
\ee
where again we have only kept terms to lowest order in $1/M^4$.
The conservation equation is given by
\be
\dot\rho+3H (\rho+p)=\frac{2}{M^4} \dot\phi\ddot\phi\rho\;,
\ee
which has solution to leading order in $1/M^4$
\be
\rho
 = \left(1+\frac{\dot\phi^2}{M^4}\right)\frac{\rho_0}{a^{3(1+w)}}\;,
\ee
for a fluid of equation of state $w$.
This follows directly from the fact that the energy densities in the Einstein and Jordan frames are related by
$
\rho=\frac{\sqrt{-\tilde g}}{\sqrt{-g}} \tilde T^{00}
$
where the conserved energy  is in the Jordan frame and therefore
$
\tilde T^{0}_0= -\frac{\rho_0}{a^{3(1+w)}}$,  $ \tilde T^{00}=\tilde g^{00} \tilde T^0_0$
while the ratio of the determinants is
$
\frac{\sqrt{-\tilde g}}{\sqrt{-g}}=(1-\frac{2\dot\phi^2}{M^4})^{1/2}.
$
Similar equations have been studied numerically in
\cite{vandeBruck:2013yxa} to all order in $1/M^4$ and where only radiation is conformally coupled.
To leading order in $1/M^4$,
the Klein-Gordon equation then becomes
\be
(\ddot \phi + 3H \dot \phi) \left(1-\frac{2p}{M^4}\right) +\frac{\partial V}{\partial \phi}=0\;.
\label{eq:KGtoM4}
\ee
In this work we will mainly consider the matter dominated era of the Universe's evolution,  where $p\approx 0$, implying that one can
restrict oneself to solving the usual Klein-Gordon equation at the background level.

\section{Electrodynamics with a disformal coupling}
Our focus in this work is on possible imprints of the disformal coupling on cosmological observations. As the vast majority of these observations are performed with light, in this Section we look in detail at how a disformally coupled scalar field interacts with the photon.

At this point in the discussion we also generalise the situation slightly by introducing a field dependent coupling constant, controlled by a new unknown scale $\Lambda$, so that the kinetic term for photons contains
\be
S_{\rm rad}\supset -\int d^4 x \sqrt{-\tilde g} \frac{1}{4}\left(1+\frac{4\phi}{\Lambda}\right) F^2\;,
\ee
where contractions are made with the Jordan frame metric. This interaction between the scalar field and photons is similar to that of an axion,
meaning that our scalar field also behaves as an axion-like particle  (ALP).
In particular, the fine structure constant becomes field dependent
\be
\alpha (\phi)= \frac{\alpha_\star}{1+\frac{4\phi}{\Lambda}}
\ee
where $\alpha_\star$ is its value in the absence of coupling.

To leading order in $1/M^4$ the photon  Lagrangian becomes
\be
{\cal L}= \sqrt{-g}\left( -\frac{1}{4} F^2 -\frac{\phi}{\Lambda} F^2 + \frac{1}{M^4} \partial_\mu\phi\partial_\nu\phi T_{(\gamma)}^{\mu\nu}\right)\;,
\label{eq:photlag}
\ee
where $ T_{(\gamma)}^{\mu\nu}= F^{\mu\alpha}{F^\nu_\alpha} -\frac{g^{\mu\nu}}{4} F^2 $ is the Einstein frame energy-momentum tensor of the photon.

By making the photon coupling constant scalar field dependent we modify the Jordan frame conservation equation.  The photon  energy-momentum tensor in the Jordan frame is
\be
\tilde T^{(\gamma)}_{\mu\nu}=F_\mu^a F_{\nu a} -\frac{\tilde g_{\mu\nu}}{4}F^2\;,
\ee
where contractions are made using the Jordan frame metric $\tilde{g}_{\mu\nu}$ given in Equation (\ref{eq:tildemetric}).
Diffeomorphism invariance implies that
\be
\tilde D_\mu \left[\left(1+\frac{4\phi}{\Lambda}\right) \tilde T^{\mu\nu}_{(\gamma)}\right]=0\;,
\ee
 and therefore
\be
\tilde D_\mu \tilde T^{\mu\nu}_{(\gamma)}= -\frac{4\partial_\mu \phi}{\Lambda} \tilde T^{\mu\nu}_{(\gamma)}\;.
\ee
We see that making the photon coupling constant (and therefore the fine-structure constant) scalar field dependent means that  the Jordan frame  photon energy-momentum tensor is not conserved any more.

\subsection{Maxwell's equation}
The equation of motion resulting from the Lagrangian in Equation (\ref{eq:photlag}) gives the generalised form of Maxwell's equation
 \be
 \partial_\alpha \left[ \left(1+ \frac{4\phi}{\Lambda} +\frac{1}{M^4} (\partial\phi)^2\right)F^{\alpha \beta}\right] -\frac{2}{M^4} \partial_\alpha\left[\partial^\mu\phi\left(\partial^\alpha \phi F_\mu^\beta -\partial^\beta\phi  F_\mu^\alpha\right)\right]=0\;.
 \ee
Notice that both terms are odd under $\alpha\to\beta$. Considering the background cosmological evolution, in which the scalar field varies in time but not with spatial position, and working in the  conformal Lorentz gauge where  $\partial_\alpha A^\alpha=0$ and $A_0=0$ we find
 \be
 -\partial_0[ C^2(\phi, \phi^{\prime}) \partial_0 A^i]+ D^2(\phi, \phi^{\prime}) \Delta A^i=0\;,
 \ee
 where $\Delta = \partial_i\partial_i$, the index $i$ runs only over spatial directions, and 
 \be
 C^2(\phi, \phi^{\prime})= 1+\frac{4\phi}{\Lambda} +\frac{1}{M^4 a^2} \phi'^2,\ D^2(\phi, \phi^{\prime})= 1+\frac{4\phi}{\Lambda} -\frac{1}{M^4 a^2} \phi'^2\;,
\label{eq:CandD}
\ee
 where $'=\partial_0$ is the derivative in conformal time $\eta$ with $ds^2=a^2(-d\eta^2 +dx^2)$.

Defining a new canonically normalised vector field as $ A^i=C^{-1} a^i$ we find that
in Fourier space
 \be
 \partial_0^2 a^i + (c_p^2 k^2 -C^{-1} C'') a^i=0\;,
\label{eq:cannormwaveeqn}
 \ee
where  the effective speed of light is $ c_p= D(\phi, \phi^{\prime})/C(\phi, \phi^{\prime})$. If $C$ and $D$ are close to one\footnote{We assume this to be the case because otherwise scalar loop corrections would destabilise the scalar Lagrangian that we started with.}, we find that
 \be
 c_p^2= 1-\frac{2}{M^4 a^2} \phi'^2\label{eq:cp}\;.
 \ee
This is expected   as the metric $\tilde g_{\mu\nu}$ in the Jordan frame is the one directing the photon trajectories.

\subsection{Eikonal approximation}
In order to see how the disformal coupling to scalars affects cosmological observations  we need to go from the generalisation  of Maxwell's equation to understanding the propagation of light rays.
When the functions $C(\phi, \phi^{\prime})$ and $D(\phi, \phi^{\prime})$, defined in equation (\ref{eq:CandD}),  vary over cosmological times, we expect that  $C''/C \propto H^2$, and similarly for $D$ (although there may be specific models in which the  variation of $C$ and $D$ is more abrupt, and these must be treated separately from this general case). Then in the sub-horizon limit $k/a\gg H$, we can neglect the effect of $C''$ in equation (\ref{eq:cannormwaveeqn}) and write the time dependent dispersion relation as
\be
\omega^2= c_p^2(\eta) k^2\;.
\ee
Let us define
\be
A_i= e_i A \cos \left(k \left[\int c_p d\eta\right] -kx+\varphi_0\right)\;,
\label{eq:Aprop}
\ee
where $A$ is the  amplitude of the photon, whose  time variation we assume to be negligible compared to the variation of the phase, which we write as
\be
\varphi=k \left(\int c_p d\eta\right) -kx +\varphi_0\;.
\ee
The polarisation vector $e^i$,  satisfies  $e^ik_i=0$ and $e^2=1$.
Using this solution of Maxwell's equation, we will calculate the energy momentum tensor in the Jordan frame.

To do this it is convenient to use conformal coordinates for the metric in the $\tilde g_{\mu\nu}$ and write
\be
d\tilde s^2= a^2(-d\tilde \eta^2 + dx^2)\;,
\ee
where $d\tilde \eta= c_p d\eta$ corresponding to the Jordan frame metric $\tilde g_{\mu\nu}= {\rm diag}(-a^2,a^2,a^2,a^2)$.
The wave function becomes simply
\be
A_i= e_i A \cos (k \tilde \eta -kx+\varphi_0)\;,
\ee
In the following we will  raise and lower indices with $\tilde g_{\mu\nu}$.

We are interested in the photon energy momentum tensor $\tilde T_{\mu\nu}^{(\gamma)}$ in the Jordan frame.
With this
we find that
\begin{equation}
\tilde T_{\mu\nu}^{(\gamma)}=\frac{A^2k^2}{a^2}\sin^2 \varphi \left[ \delta_{\mu \tilde{\eta}}\delta_{\nu \tilde{\eta}}-\frac{2k_i}{k}\delta_{\tilde{\eta} (\mu}\delta_{\nu)}^i +\frac{2k_ik_j}{k^2}\delta_{(\mu}^i\delta_{\nu)}^j\right]\;,
\end{equation}
where the round brackets denote symetrization over the indices.
Changing back to the true conformal time coordinate $\eta$, whose directions we label with $0$ and using the fact that $\frac{d\tilde\eta}{d\eta}=c_p$ we find that
\begin{equation}
\tilde T_{\mu\nu}^{(\gamma)}=\frac{A^2k^2}{a^2}\sin^2 \varphi \left[c_p^2 \delta_{\mu 0}\delta_{\nu 0}-\frac{2c_p k_i}{k}\delta_{0 (\mu}\delta_{\nu)}^i +\frac{2k_ik_j}{k^2}\delta_{(\mu}^i\delta_{\nu)}^j\right]\;.
\end{equation}
This can be written in the form of the
energy-momentum tensor typically used in  geometrical optics
\be
\tilde T_{\mu\nu}^{(\gamma)}= \frac{A^2} {a^2} k_{\mu}k_\nu\;,
\label{eq:energymomentumgeometrical}
\ee
where $k_\mu=(-c_p k \sin (\varphi),k_i \sin (\varphi))$, and  $k_\mu=\partial_\mu g$,
where $g= \cos (\varphi)$.

Notice that
\be
\tilde T^{(\gamma)0}_0= -\frac{A^2 k^2}{2a^4}=-\tilde \rho_\gamma\;,
\ee
is the radiation energy density. The conservation equation for the radiation fluid in the Jordan frame becomes
\be
\dot {\tilde \rho}_{(\gamma)}+ 4H \tilde \rho_{(\gamma)}=-\frac{4}{\Lambda}\dot \phi\tilde \rho_{(\gamma)}\;,
\ee
implying that
\be
A^2={A_0^2} e^{-\frac{4}{\Lambda}(\phi-\phi_0)}=
{A_0^2} \frac{\alpha}{\alpha_0}\;,
\label{con}
\ee
to leading order in $\phi/\Lambda$, where $A_0$, $\phi_0$ and $\alpha_0$ are constants of integration.
Therefore the photon intensity evolves only when the fine structure constant varies.  In particular, we find that the intensity only varies along photon trajectories if the fine structure constant varies, driven by the scalar field evolution.  This is an effect which could be seen in the CMB spectrum, and was first studied in the context of axions and axion-like particles in \cite{Mirizzi:2005ng}.

\subsection{Geodesics}
 At this point we know that even when the speed of light varies, Maxwell's equation leads, in the geometrical optics approximation $\partial_0 \varphi \gg \partial_0 A$, to the propagation
of light rays. We now need to confirm that  these light rays follow geodesics.
Notice first of all that
\be
\tilde g^{\mu\nu} k_\mu k_\nu=0\;,
\ee
showing that $k_\mu$ is null for $\tilde g_{\mu\nu}$.
Acting with the covariant derivative defined with respect to $\tilde g_{\mu\nu}$, we have
\be
(\tilde D_\lambda k_\nu) k_\mu \tilde g^{\mu\nu}=0\;,
\ee
and upon using
$ \tilde D_\lambda k_\nu= \tilde D_\lambda \partial_\nu g= \tilde D_\nu\partial_\lambda g= \tilde D_\nu k_\lambda$, we have
\be
\tilde g^{\mu\nu}k_\mu\tilde D_\nu k_\lambda=0\;,
\ee
which is the geodesic equation.  Therefore we conclude that  light rays follow geodesics of $\tilde g_{\mu\nu}$.
Along these geodesics the speed of light varies, which is apparent in the phase of the solution to Maxwell's equation. This has consequences for the way of measuring distances in cosmology.

\section{Distance Duality Relations}
If photons are no longer conserved and the paths on which they travel are modified by the disformal coupling then our understanding of distance measures in cosmology will be modified.  In this section we briefly review the standard relationships between different distance measures  and show how these are modified by the couplings to the scalar field that we have introduced.

There are two types of distances that can be inferred from observations that are commonly used in cosmography.  The angular diameter distance $d_A$ of an object is obtained by considering a bundle of geodesics converging at the observer under a solid angle $d\Omega_{\rm obs}$ and coming from a surface area $dS_{\rm emit}$:
\be
d_A^2= \frac{dS_{\rm emit}}{d\Omega_{\rm obs}}\label{eq:areadistance}\;.
\ee
The luminosity distance is given in terms of the emitter luminosity $L_{\rm emit}$ and the radiation flux received by the observer $F_{\rm obs}$ by
\be
d_L^2= \frac{L_{\rm emit}}{4\pi F_{\rm obs}}\;.
\label{eq:lumdist}
\ee
For a unit sphere,
\be
L_{\rm emit}= \int F_{\rm emit} d\Omega_{\rm emit}= 4\pi F_{\rm emit}\;,
\ee
where $F_{\rm emit}$ is the emitted  flux.

The luminosity and angular distances are related in the standard cosmology by Etherington's theorem\cite{Ellis:1971pg}:
\begin{equation}
d_L(z)=(1+z)^{2} d_A(z)\;.
\end{equation}
This is also known as the distance duality relation. The validity of the distance duality relation requires that photons propagate along null geodesics and that the geodesic deviation equation holds.  In addition, the number of photons must be conserved.
If this last hypothesis fails, then we can still find an expression relating the area distances up and down the past light cone. This is known as the reciprocity relation.
  Let $r_S$ be the source area distance, which we define by considering a bundle of null geodesics diverging from the source and subtending a  solid angle $d\Omega_S$ at the source.  The bundle has a cross section $d S_O$ at the  observer.  Then the source area distance is defined as $ d S_O=r_S^2 d\Omega_S$. This definition is complementary to the one of angular distance given above, where instead the solid angle subtended by the bundle converging  at the observer and the section of the bundle at the source are considered (eq. (\ref{eq:areadistance})). In the context of the reciprocity relation usually the convention is to call $r_{O}\equiv d_{A}$.
  The reciprocity relation (again in the standard cosmology) states
	\begin{equation}
	r_S =(1+z)r_O\;.
	\end{equation}
As mentioned above, the reciprocity relation requires only that photons propagate along null geodesics and that the geodesic deviation equation holds.  The distance duality relation requires, in addition, that the number of photons is conserved. So we expect that modification of geodesics will affect both the reciprocity and distance duality relations, while the non conservation of photons will only affect the distance duality relation.

We now proceed to show how these relations are modified in the presence of couplings between the scalar field and photons.   In this discussion we will use the notation employed in recent work by Ellis, Poltis, Uzan and Weltman \cite{Ellis:2013cu} in order to allow for comparison with that work.

Firstly we need to  state carefully what we mean by redshift.
Along one geodesic, if we  define $u^\mu$ to be  the velocity of an observer with $u_{\mu}u^{\mu}=-1$, the frequency of a photon is
\be
E= -k_{\mu}u^{\mu}\;.
\ee
This is expected for photons with a modified dispersion relation in a cosmological background, as can be seen by going to the rest frame of a comoving observer  where $u^{\mu}=(a^{-1},0,0,0)$ and we find explicitly that the dispersion relation is $E=c_p k/a$ as we have found from Maxwell's equation.
We are thus able to  identify the redshift along a light ray between the emitter and the observer
\be
(1+z)= \frac{(k_{\mu}u^{\mu})_{\rm emit}}{(k_{\mu}u^{\mu})_{\rm obs}} \;.
\ee

Now  using the fact that light rays are geodesics and the geodesic deviation equation
(that says that geodesics deviate according to the curvature tensor), we have that\cite{Ellis:1971pg}
\be
\frac{dS_{\rm emit}}{d\Omega_{\rm obs}}= \frac{(k_a\tilde u^a)^2_{\rm obs}}{(k_a\tilde u^a)^2_{\rm emit}} \frac{dS_{\rm obs}}{d\Omega_{\rm emit}}\;,\label{eq:dSdOmega}
\ee
where $\tilde g_{\mu\nu} \tilde u^\mu \tilde u^\nu=-1$ and $\tilde u^\mu$ is the velocity vector of a comoving observer on the geodesics followed by the photons.
Now for comoving observers  and emitters  in their  rest frame we have
$\tilde u^\mu= (a^{-1} c_{p}^{-1},0,0,0)$ and
we find that $k_a \tilde u^a= k/a$ and $k_a u^a= c_p k/a$, implying that $k_a \tilde u^a=c_p^{-1} k_au^a$, so that:
\be
\frac{(k^a\tilde u_a)_{\rm emit}}{(k^a \tilde u_a)_{\rm obs}}= \frac{c_{\rm obs}}{c_{\rm emit}} (1+z)\;.\label{eq:betadef1}
\ee
Considering again the source area distance, defined above as
\be
r_{S}^2= \frac{dS_{\rm obs}}{d\Omega_{\rm emit}}\;,
\ee
and comparing it to the  angular diameter distance (\ref{eq:areadistance}),
we find through equations (\ref{eq:dSdOmega}) and (\ref{eq:betadef1}) that the violation of the reciprocity relation is described by the function $\beta $
\be
r_S^2=  (1+z)^2 \beta(\eta_{\rm obs},\eta_{\rm emit}) d_A^2\;,
\ee
where $\beta$ is defined as:
\be
\beta(\eta_{\rm obs},\eta_{\rm emit})=
\left(\frac{c_{\rm obs}}{c_{\rm emit}}\right)^2\;.
\ee
So, as expected, we see that the reciprocity relation is modified if photons follow modified geodesics.

To understand how the distance duality relation is violated due to non-conservation of photons along geodesics, we use the conservation of  $(1+4\frac{\phi}{\Lambda}) a^2A^2 dS$  along a bundle of light rays.
This is true even in the presence of a disformal scalar field, as a consequence of the conservation of $(1+4\frac{\phi}{\Lambda})\tilde T_{\mu\nu}$.  Therefore  along light rays
\be
k^\mu \tilde D_{\mu} \left[\left(1+4\frac{\phi}{\Lambda}\right)a^2A^2\right]=- \left(1+4\frac{\phi}{\Lambda}\right)a^2A^2 \tilde D_\mu k^\mu\;,
\ee
where again $A$ is the amplitude of the propagating photon wave.  The following result, which is valid for any null bundle (see chapter 11 in \cite{lud})
\be
k^\mu \tilde D_\mu S=( \tilde D^\mu k_\mu) S\;,
\ee
implies that
\be
k^\mu \tilde D_\mu \left(\left[1+4\frac{\phi}{\Lambda}\right]a^2A^2 S\right)=0\;,
\ee
meaning that the product $(1+4\frac{\phi}{\Lambda})a^2A^2 dS$ is conserved along light rays.

The luminosity of a source is defined as the energy flux across the surface area $dS$
\be
dL= a^2A^2 (k^a u_a)^2 dS\;,
\ee
and the flux of light from the source is
\be
F= a^2A^2 (k^a u_a)^2\;.
\ee
From the conservation of $(1+4\frac{\phi}{\Lambda})a^2A^2 dS$, and using the fact that $dS=d\Omega$ for a sphere of unit radius, we have that at the observer
\be
F_{\rm obs}=\frac{\alpha_{\rm obs}}{\alpha_{\rm emit}}\frac{F_{\rm emit} d\Omega_{\rm emit}}{(1+z)^2 dS_{\rm obs}}\;,
\ee
where $F_{\rm emit}$ is the flux on a unit sphere around the emitter.
Now from the definition of the luminosity distance in equation (\ref{eq:lumdist})  we find that
\be
d_L^2= (1+z)^2 \frac{\alpha_{\rm emit}}{\alpha_{\rm obs}}\frac{dS_{\rm obs}}{d\Omega_{\rm emit}}\;.
\label{eq:luminositydistance}
\ee

The duality relation\footnote{We have implicitly averaged over $\varphi_0$ to define the flux and the luminosity.} is modified by a function $\tau$
\be
d_L= \tau(\eta_{\rm obs},\eta_{\rm emit})  (1+z)^2 d_A\;,
\label{eq:dLdAtau}
\ee
where
\be
\tau(\eta_{\rm obs},\eta_{\rm emit})^2 = \left(\frac{\alpha_{\rm emit}}{\alpha_{\rm obs}}\right)\beta(\eta_{\rm obs},\eta_{\rm emit})\;.
\label{eq:tau}
\ee
Notice that the duality relation is modified due to two different effects. First the flux is reduced by the variation of $\alpha$ and then the geodesics are affected by the change of the speed of light. The first change appears in the luminosity distance and the second one in the reciprocity relation.  If both the luminosity distance and the angular diameter distance can be measured as a function of redshift, then a lack of violation of the duality relation can be used to constrain the interactions of the disformal scalar field with photons.

\section{CMB Distortions}

The interaction of the disformal scalar field with photons has implications for any observations done using photons, and precision measurements  allow us to use these observations to constrain the influence of such scalar fields.    The most precise cosmological observations that we have to date are those of the cosmic microwave background (CMB) and in this Section we analyse how this may be distorted by the presence of a disformal scalar.

\subsection{Photon intensity}
We return to the expression for the propagating photon derived in Equation (\ref{eq:Aprop}). In the Lorentz gauge and assuming that propagation is along the z axis, we have $A^z=0$.
 As long as $C(\phi,\phi^{\prime})$ and $D(\phi,\phi^{\prime})$ are slowly varying functions of $\eta$, we have the WKB solution
 \be
 A^{x,y}(k,\eta)=A^{x,y}(k,\eta_i)\sqrt{\frac{\omega(k,\eta_i)}{\omega(k,\eta)}}\frac{C(\eta_i)}{C(\eta)} \cos \left(\int_{\eta_i}^\eta \omega(k,\eta') d\eta' -kz +\varphi_0\right)\;,
 \ee
 where $\varphi_0$ is an initially random phase, and $\eta_i$ is the initial conformal time.
 We have defined
 \be
 \omega^2(k,\eta)= c_p^2 k^2 -C^{-1} C''\;.
 \ee
 The intensity of the photon radiation in the conformal gauge is given $I= a^4 \rho_\gamma$ and therefore
 \be
 I=\frac{1}{2} [(\partial_0 A^i)^2 + B_iB^i]\;,
 \ee
 where indices are raised with $\eta_{\mu\nu}$ and  $B_i=\epsilon_{ijk} \partial^j A^k$ .
 As long as $C$ and $D$ vary slowly we find
 \begin{eqnarray}
 I(k,\eta) &=& \frac{1}{2} \sin^2\left(\int_{\eta_i}^\eta \omega(k,\eta') d\eta' -kz +\phi_0\right)\left(k^2 +\omega^2 (k,\eta)\right) \\
& & \times \left[(A^x(\eta_i,k))^2+(A^y(\eta_i,k))^2\right]{\frac{\omega(k,\eta_i)}{\omega(k,\eta)}}\frac{C^2(\eta_i)}{C^2(\eta)}\;. \nonumber
 \end{eqnarray}
 Averaging over $\varphi_0$ gives
 \be
 I(k,\eta)= G(k,\eta,\eta_i) I(k,\eta_i)\;,
\label{eq:Iketa}
 \ee
 where
 \be
 G(k,\eta,\eta_i)={\frac{\omega(k,\eta_i)}{\omega(k,\eta)}}\frac{C^2(\eta_i)}{C^2(\eta)}\frac{k^2 +\omega^2 (k,\eta)}{k^2 +\omega^2 (k,\eta_i)}\;.
 \ee
We find that the scalar field induces a time variation in the frequency of the photons and also induces directly a time variation in the intensity.

\subsection{$\mu$ distortion}
Our understanding of the primordial Universe leads us to expect that the CMB will display an almost perfect black body spectrum.    Any new physics that interferes with photons, particularly in a frequency dependent way will distort the black body spectrum of the CMB and so can be constrained by these observations.

We consider that the CMB spectrum is initially a black body spectrum, so that
\be
I(k,\eta_i)= \frac{k^3}{e^{k/T_0}-1}\;,
\ee
and we assume that the only distortions appear through the influence of the scalar field as the light propagates towards us from the time of last scattering.  The measured spectrum will be related to the intensity $I(k,\eta)$ of equation (\ref{eq:Iketa}) by a geometrical factor which depends on the way the reciprocity relation is modified by the variation of the speed of light \cite{Ellis:2013cu}
\be
I_{\rm obs}(k,\eta)= \left(\frac{d_A}{r_{\rm emit}}\right)^2 G(k,\eta,\eta_i) I(k,\eta_i)\;.
\ee
The first factor depends on the ratio of the emitted to the observed speeds of light, the second factor $G$ appears because of the attenuation of the amplitude due to the change of the fine structure constant and the exchange of energy with the scalar field. The combined effect is therefore
\be
I_{\rm obs}(k,\eta)=  \tau^{-2}(\eta,\eta_i)I(k,\eta_i)\;,
\ee
where the function $\tau$ was defined in equation (\ref{eq:tau}), a result which is valid on subhorizon scales and when the speed of light varies cosmologically.  Including  the time evolution of $\phi$ we can rewrite the intensity  as
\be
I_{\rm obs}(k,\eta)= \frac{k^3}{e^{k/T_0+\mu}-1}\;,
\ee
where the adimensional chemical potential is given by
\be
\mu=  -2(e^{-k/T_0}-1)\delta \tau \;,
\label{eq:mu}
\ee
where
$
\tau(\eta,\eta_i)=1 + \delta \tau $,
and
\be
\delta \tau = \frac{1}{2} \frac{\delta \alpha}{\alpha} -\frac{\delta c_p}{ c_p}\;,
\label{eq:deltaGdeltaphi}
\ee
where $\delta \alpha$ and $\delta c_p$ denote the difference in these quantities between their current and their initial values.
Hence we have found that the CMB spectrum and the duality relation are distorted due to the same $\tau$ function whose origin follows from both the disformal coupling and direct coupling of a scalar field to
photons. We will constrain these couplings in the following section.

\section{Observational constraints}

The time variations of the fine structure constant  and the speed of light depend on the dynamics of $\phi$. In this Section we use observations of the CMB and of the distance duality relation to constrain these variations. The scalar field's dynamics are not affected by its coupling to $F^2$ as on average for a radiation fluid we have $\langle F^2\rangle= 0$ (as the electric and magnetic field have the same amplitudes). Hence in the matter dominated era the equation of motion for the scalar field given in (\ref{eq:KGtoM4}) becomes
\be
\ddot \phi + 3H \dot \phi+\frac{\partial V}{\partial \phi}=0\;.
\ee
 The evolution of the scalar field is determined entirely by the scalar field potential (and the background Hubble evolution). In Section \ref{sub:constraintsThawing} we will focus on an example of how these constraints restrict the parameters for a technically natural choice for the potential.

\subsection{Variations of $\alpha$}

 As we have seen observations of CMB spectral distortions, and of the distance duality relation both constrain the same combination of the variation in $\alpha$ and the variation in the speed of light. In principle, by using observations of the CMB at high redshift and observations of the distance duality relation at comparatively small redshift we can break the parameter degeneracy.  However as we will see in the next Section, in models in which the scalar field does not roll at early times, the two types of cosmological observation discussed here constrain the same combination of parameters, because the scalar field has not evolved between the times of these two classes of observations.  Breaking this degeneracy would be possible if observations could be made before and after the field starts rolling.

Independent constraint on either $\delta\alpha$ or $\delta c_{p}$ would allow us to give independent estimates on the two variations and then on $\Lambda$ and $M^{4}$.
In Table \ref{tab:deltaalphaconstraints} we report a summary of presently available constraints on $\delta\alpha$. These observations are independent of the CMB distortions and distance duality violations caused by the disformal scalar field, and it is difficult to see how the presence of a disformal scalar coupling could cause them to be misinterpreted.  We work under the optimistic assumption that the reported constraints on variation of the fine structure constant  exclusively constrain the variable we have called  $\delta\alpha$ and not $\delta c$.  These constraints come from a variety of observations, each of them referring to variation over a different redshift range. Therefore when combining them with observations of CMB distortions or of distance duality relations our interest is focused on  constraints that refer to the same redshift range as the one we are probing.

\begin{table}[htdp]
\centering

\begin{tabular}{|c|c|c|c|}
\hline
Observation&redshift ($z$)&$(\alpha(z)-\alpha_{0})/\alpha_{0}\pm \sigma $&Reference\\
\hline
CMB anisotropies&$10^{3}$&$-0.013\pm 0.012$&\cite{Menegoni:2009rg}\\
Quasar doublet absorption lines&$2\sim4$&$(-4.6\pm4.5)\cdot 10^{-5}$&\cite{Varshalovich:2000sb}\\
Quasar doublet absorption lines&$0.5\sim1.8$ ($\langle z\rangle =1$)&$(-0.70\pm0.23)\cdot 10^{-5}$&\cite{Murphy:2000pz}\\
Quasar doublet absorption lines&$0.9\sim3.5$ ($\langle z\rangle =2.1$)&$(-0.76\pm0.28)\cdot 10^{-5}$&\cite{Murphy:2000pz}\\
BBN&$10^{10}$&$(-7\pm5)\cdot 10^{-3}$&\cite{Avelino:2001nr}\\
\hline
\end{tabular}
\caption{Constraints on $\delta\alpha/\alpha\equiv \frac{\alpha(z)-\alpha_{0}}{\alpha_{0}}$. $\alpha_{0}$ is the present value of the fine structure constant}
\label{tab:deltaalphaconstraints}
\end{table}%

\subsection{$\mu$ distortion constraints}
The present limits on the amount of $\mu$ distortion in the CMB spectrum come from COBE/FIRAS observations. At $95 \%\, \text{c.l.}$ they are $|\mu|< 3.3\times 10^{-4}$ at wavelengths of $\text{cm}$ and $\text{dm}$ \cite{Firas}.
The ARCADE2 balloon  also provided constraints on $\mu$ spectral distortion, $|\mu|<6\cdot 10^{-4}$ at $95\,\%$ c.l. between $3$ and $90$ GHz \cite{ARCADE2}.
The proposed experiment PIXIE \cite{PIXIE} will be sensitive to $\mu\sim 10^{-8}$.
In this Section we will use the current observations of the spectral distortions to  put limits on $\delta \tau$, defined in equation  (\ref{eq:deltaGdeltaphi}).  We will also discuss the power of future probes to constrain these models.

 We  assume that the constraining power of observations of the black body spectrum of the CMB comes from observations at frequencies corresponding to $T_{0}\sim 2.7 K$ (see Fig. 12 of \cite{PIXIE} for a discussion of why this is reasonable). Therefore we find
\be
|\mu|< 3.3\times 10^{-4} \Rightarrow |\delta \tau|< 2.6\times 10^{-4}\;.
\label{eq:deltaalphafrommu}
\ee
The severity of  the constraints on the scalar field  depends on the frequency of light at which the constraints are applied.  If we were to assume that constraints on spectral distortions of the CMB of this strength were able to be extended to frequencies as low as $30$ GHz, then we would get the weaker constraint $|\delta \tau|<9.5 \cdot 10^{-4}$, while if we assume that the spectral distortion constraint can be extended to frequencies as high as $300$ GHz, then we would infer $|\delta \tau|<2.0\cdot 10^{-4}$. These are the two limiting cases as the CMB spectrum is not measured well outside this range.

The limit given in equation (\ref{eq:deltaalphafrommu}) constraints a combination of the fine structure constant and speed of light variation, and hence a combination of the coupling constants $\Lambda$ and $M^{4}$ (once the scalar field potential is fixed). To break this degeneracy we use constraints on the variation of $\alpha$.  The measurement of interest is that  reported in the first line of  Table \ref{tab:deltaalphaconstraints}.  We stress that although these constraints come from observations of the CMB,  they  come from observations of the properties of the fluctuations  and therefore can be considered as observables that are independent of the CMB distortions.  Unfortunately, since the constraint on $\delta\alpha/\alpha$ there reported is much weaker than the one we have on the combined quantity $\frac{1}{2}\frac{\delta\alpha}{\alpha}-\frac{\delta c_{p}}{c_{p}}$, we cannot draw any conclusion about $\delta c_{p}/c_{p}$.

Concerning improvements to be expected from future probes of CMB spectral distortion, notice that $\mu$ is linearly dependent on $\delta\tau$. So the four-orders-of-magnitude improvement expected from PIXIE would translate into a constraint on $|\delta\tau|$ at the level of $10^{-8}$.

\subsection{Distance Duality constraints}

We emphasize that we need  distance duality relation constraints to be as independent as possible from  the cosmological model, so that we can use them in our framework,  without worrying about the effects of the scalar field on the evolution of the Universe  or loose constraining power by having to take the variation of other variables into account.
The best current  constraint of this kind is provided in reference \cite{Holanda:2012at}.
There we find a comparison between galaxy cluster mass fraction estimations  obtained from X-ray measurements (which probe $d_{L}/d_{A}$) and observations of the Sunyaev-Zeldovic effect  (which probe $d_{A}$). The clusters  considered are all in the redshift range $z\in(0.1,0.9)$.  We assume that the galaxy cluster mass estimate obtained from the X-ray observations is not influenced by the effects of the disformal scalar field, we will find that quasar observations will constrain the variations of the fine structure constant induced by the scalar field to be  small, and we can check therefore that any mixing between the scalar field and photons in the cluster magnetic fields is negligible, justifying our  assumption.

In an unfortunate clash of notations, in  \cite{Holanda:2012at} the violation of the distance duality relation is parametrised by  $\beta(z)$, defined by $d_L= \beta(z)  (1+z)^2 d_A$, which is then equivalent to our function $\tau$ in  equation (\ref{eq:dLdAtau}); if we define  $\tau(\eta_{obs},\eta_{emit})\equiv\tau(z)$; where we put $z_{obs}=0$ and $z_{emit}=z$. A specific functional form for $\beta$ is assumed in \cite{Holanda:2012at} ,  $\beta(z)=1+\beta_{0}z$, which we can also consider valid in the small redshift range probed by observations, since in that case we can expand $\tau(z)$ at the first order in $z$ as $\tau(z)=1+\frac{d\tau}{dz}|_{z=0}\,\,z$. So the constraints on $\beta_{0}$ reported in  \cite{Holanda:2012at} are equivalent to constraints on $\frac{d\tau}{dz}|_{z=0}$ in our more general case.

The constraint found in \cite{Holanda:2012at} is:
\begin{equation}
\beta_{0}=-0.06\pm0.16\;,
\end{equation}
at the 2$\sigma$ level.
We  translate this into a constraint at an intermediate value of the redshift $\tau(z=0.35)=0.979\pm0.056$ at 2$\sigma$, which can also be stated as $\delta\tau(z=0.35)=-0.021\pm0.056$.


We can combine this estimate with the one on $\frac{\delta\alpha}{\alpha}$ coming from Quasar observations, in the third line of Table \ref{tab:deltaalphaconstraints}, which probes a similar redshift range, in order to get an estimate on $\frac{\delta c_{p}}{c_{p}}$. In this case, since the constraint on  $\frac{\delta\alpha}{\alpha}$ by itself is so much tighter than the one we get from the distance duality relation, we can just assume that all the contribution to $\delta\tau$ comes from the speed of light variation:

\begin{equation}
\frac{\delta c_{p}}{c_{p}}=0.021\pm0.056\;,
\end{equation}
or:

\begin{equation}
\left|\frac{\delta c_{p}}{c_{p}}\right|<0.060 \label{eq:deltacconstraint}\;,
\end{equation}
at $68\%$ c.l. assuming a gaussian distribution of errors.

\section{Constraints on Axionic quintessence models}
\label{sub:constraintsThawing}

As discussed in the Introduction disformal couplings will naturally arise in dark energy models which possess an axionic shift symmetry, such models have also been termed `thawing quintessence' because of the dynamics of the scalar field.  They are typically described by a scalar field with potential
\begin{equation}
V(\phi) =\frac{\Lambda_0^4}{2}\left(1+\cos \frac{\phi}{f}\right)\;,
\end{equation}
where $\Lambda_0$ and $f$ are constant model parameters.  It is assumed that initially $\phi_i \ll f$ and that the field only starts rolling at late times in the history of the universe.  This model is described in detail in Appendix \ref{sec:app}.

We derive constraints on the axionic quintessence model  using the information on variation of $\alpha$ and variation of the speed of light as derived in the previous Section. The constraint on  $\delta c_{p}/c_{p}$ refers to variation between redshift $z_{e}\sim0.35$ (average redshift of measured clusters) and now. The constraint on  $\delta \alpha/\alpha$ refers to variation between redshift $z_{e}\sim 1$ (average redshift of measured quasars, see Table \ref{tab:deltaalphaconstraints}, third line) and now.
In Appendix \ref{sec:app} we show that slow rolling starts at redshift $z_{r}$ such that $z_{r}<1$.
So when translating the  constraint on $\delta\alpha/\alpha$ into a constraint on the quintessence model we always have $z_{r}\le z_{e}$, while for the case of $\delta c_{p}/c_{p}$ we have two cases: $z_{r}< z_{e}$ and $z_{r}\ge z_{e}$.

It is straightforward to find expressions for the variations in fine structure constant and the speed of light as a function of the scalar field parameters.  The details of this calculation are given in the Appendix, and we just quote the results here for $z<z_{r}$:
\be
\frac{\alpha}{\alpha_\star}= 1 -4\sqrt{6} \sqrt{1+w_\phi}\frac{ f}{m_{\rm Pl}}\frac{m_{\rm Pl}}{\Lambda} \left(1-\frac{z}{ z_r}\right)\;,
\label{eq:deltaalphaf1}
\ee
\be
c_p^2 =1 -\frac{16(1+w_\phi)}{3 } \frac{H_0^2 m_{\rm Pl}^2}{M^4} \frac{f^2}{m_{\rm Pl}^2}\left(1-\frac{z}{ z_r}\right)^2 \;.
\label{eq:deltacf1}
\ee

\begin{figure}[h!]
\centering
\fbox{\includegraphics[scale=1]{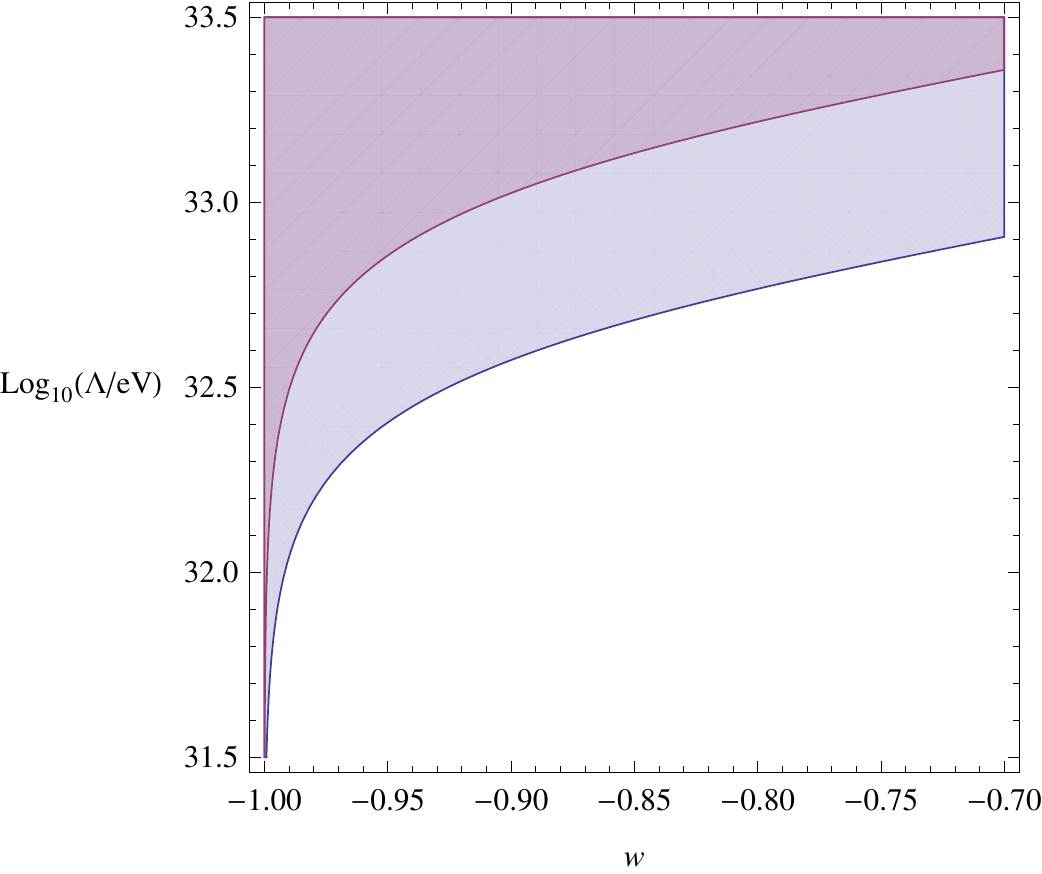}}
\caption{\footnotesize The coloured region is the one allowed for $\log_{10}(\Lambda/eV)$, the white one is excluded. The two colours refer to the limiting cases $z_{r}=1$ (blue region) and $z_{r}\rightarrow0$ (purple region). The figure refers to $68\%$ c.l.}
\label{fig:LambdaExclusionPlot}
\end{figure}

Let us first consider the variation of fine structure constant $\alpha$. Since $z_{r}\le z_{e}$, the field starts slow rolling after radiation is emitted, and  we can assume that  the value of  $\phi$ at the time of emission is zero (and consequently $\alpha=\alpha_\star$). After the beginning of slow rolling ($z<z_{r}$) the field evolution follows eq. (\ref{eq:phiz}) and consequently
\begin{equation}
\frac{\delta\alpha}{\alpha}=4\sqrt{6}\sqrt{1+w_{\phi}}\frac{f}{\Lambda}\;,
\label{eq:deltaalphaoveralpha}
\end{equation}
Using the constraint on $\delta\alpha/\alpha$ from Quasars observations at redshift $z\sim1$, reported in the third line of Table 1, we get:
\be
 \frac{4\sqrt{6}}{ \Lambda}\sqrt{1+w_{\phi}}f<0.81\times 10^{-5}\;,
\ee
at  $68\%$ c.l.
An expression for $f$ is reported in (\ref{eq:fsquareThawing}) in terms of the redshift at which the scalar starts rolling, the equation of state of the scalar field today, the current fraction of Dark Energy in the Universe and the Planck mass. This leads to a constraint on a combination of $\Lambda$, $z_{r}$ and the equation of state parameter $w_{\phi}$:
\be
 \frac{4\sqrt{6}}{ \Lambda}\sqrt{1+w_{\phi}}\sqrt{\frac{\frac{3}{2}\Omega_{\Lambda 0}  m_{\rm Pl}^2}{(1+z_{r})^{3}[1-\frac{3}{2}(1+w_\phi)]}}<0.81\cdot 10^{-5}\;.
\ee
Note that the dependence on $z_{r}$ is weak and can contribute only up to a factor of three \footnote{This is comparing the constraint with $z_{r}=1$ and $z_{r}\rightarrow 0$. Of course, if $z_{r}=0$, then slow roll never starts during photons propagation so that $\phi\equiv 0$ and $\delta\alpha\equiv 0$ as one can see from eq. (\ref{eq:deltaAlphazr}). Here we are considering $z_{r}\rightarrow 0$ just as a limiting case.} . We can then consider the limiting case in which $z_{r}=1$, which gives the most conservative constraint.
With  $H_{0}=2.13\cdot 10^{-33}\mbox{ eV}$, $m_{Pl}=2.44\cdot10^{27}\mbox{ eV}$, $\Omega_{\Lambda 0}=0.73$ and taking as an illustration the value of the equation of state  $w_{\phi}=-0.96$ we get:
\be
\Lambda>2.3\times 10^{32} \mbox{ eV}= 1.9 \times 10^4 m_P\;,
\ee
at $68\%$ c.l. forcing this coupling scale to lie well above the Planck scale.
In Figure \ref{fig:LambdaExclusionPlot} we show a plot with the allowed region for $\Lambda$ as a function of the equation of state parameter $w_{\phi}$, considering the two limiting cases $z_{r}=1$ (blue region) and $z_{r}\rightarrow 0$ (purple region).

The variation of the speed of light  is related to the variation of the field derivative through eq. (\ref{eq:cp}).
  If $z_{r}\ge z_{e}$ the field is already slow rolling when radiation is emitted, so that $\dot\phi\neq 0$ and it evolves according to eq. (\ref{eq:dotphiz}).
 So we have :
\begin{equation}
 \frac{\delta c_{p}}{c_{p}}=\frac{8(1+w_{\phi})H_{0}^{2}f^{2}}{3M^{4}}\frac{z_{e}}{z_{r}}\left(2-\frac{z_{e}}{z_{r}}\right)\label{eq:deltacoverc}
 \end{equation}
As before the coefficient $f$ is related to the  cosmological parameters through eq. (\ref{eq:fsquareThawing}).

\begin{figure}[h!]
\centering
\fbox{\includegraphics[scale=1]{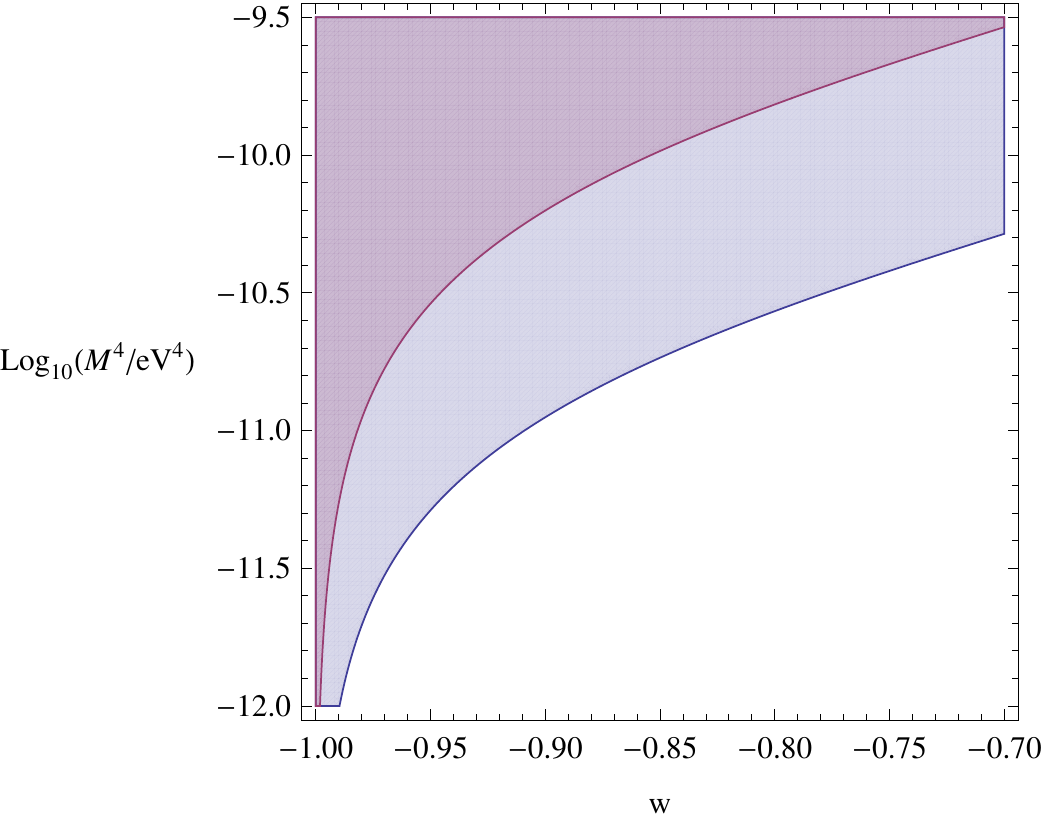}}
\caption{\footnotesize The coloured region is the one allowed for $\log_{10}(M^{4}/eV^{4})$, the white one is excluded. The two colours refer to the limiting admissible values of $z_{r}$, $z_{r}=0.35$ (purple region) and $z_{r}=1$ (blue region). The figure refers to $68\%$ c.l. This is  the case in which $z_{e}>z_{r}$.}
\label{fig:M4ExclusionPlot}
\end{figure}

Using the constraint (\ref{eq:deltacconstraint}) on $\delta c_{p}/c_{p}$  we can derive limits on a combination of $M^{4}$ and  $z_{r}$ through eq. (\ref{eq:deltacoverc}).
Using the expression for $f$ in (\ref{eq:fsquareThawing})and $z_{e}=0.35$:
\begin{equation}
\left(\frac{M}{10^{-2}\mbox{ eV}}\right)^{4}>4.6\times 10^{-2}\frac{ (1+w_{\phi})}{[1-\frac{3}{2}(1+w_{\phi})]}\frac{\left(2-\frac{0.35}{z_{r}}\right)}{{z_{r}}(1+z_{r})^{3}}\;.
\end{equation}
For an illustrative value of $w_{\phi}=-0.96$ this becomes:
\begin{equation}
\left(\frac{M}{10^{-3}\mbox{ eV}}\right)^{4}>19.5\frac{\left(2-\frac{0.35}{z_{r}}\right)}{{z_{r}}(1+z_{r})^{3}}\;.
\end{equation}
So the constraint on $M^{4}$ depends on the redshift at which the field starts slow rolling (remember that here we assumed $0.35<z_{r}<1$). Given the variability range of $z_{r}$ the constraint on $M^{4}$ can change  up to a factor of five, from $M^{4}>4\cdot 10^{-12}\,\,\text{eV}^{4}$ (if $z_{r}=1$) to $M^{4}>2\cdot 10^{-11}\,\,\text{eV}^{4}$ (if $z_{r}=0.35$).  In Figure \ref{fig:M4ExclusionPlot} we plot the allowed region for $M^{4}$ as a function of  the equation of state parameter $w_{\phi}$ for the limiting cases of $z_{r}=1$ (blue region) and $z_{r}=0.35$ (purple region).

 If $z_{r}< z_{e}$ the field is not slow rolling when radiation is emitted, so that $\dot\phi= 0$ and $\delta c_{p}(z_{e})=0$. Then for $z<z_{r}$ the field starts slow rolling and it evolves according to eq. (\ref{eq:dotphiz}).
So we have
\begin{equation}
 \frac{\delta c_{p}}{c_{p}}=\frac{8(1+w_{\phi})H_{0}^{2}f^{2}}{3M^{4}}\label{eq:deltacovercsecondcase}
 \end{equation}

\begin{figure}[h!]
\centering
\fbox{\includegraphics[scale=1]{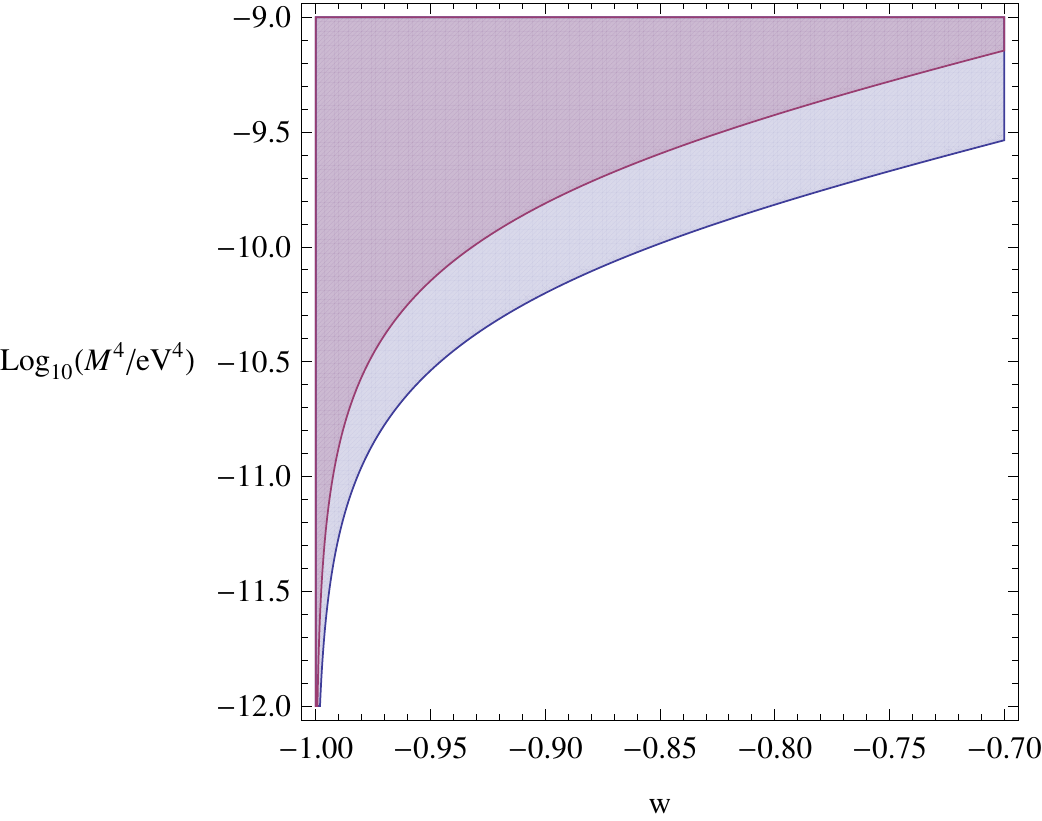}}
\caption{\footnotesize The coloured region is the one allowed for $\log_{10}(M^{4}/eV^{4})$, the white one is excluded. The two colours refer to the limiting admissible values of $z_{r}$, $z_{r}\rightarrow 0$ (purple region) and $z_{r}=0.35$ (blue region). The figure refers to $68\%$ c.l. This is  the case in which $z_{e}<z_{r}$.}
\label{fig:M4ExclusionPlotsecondcase}
\end{figure}

Using the expression for $f$ in (\ref{eq:fsquareThawing}),  the constraint (\ref{eq:deltacconstraint}) translates into:
\begin{equation}
\left(\frac{M}{10^{-2}\mbox{ eV}}\right)^{4}>0.13\frac{ (1+w_{\phi})}{(1-\frac{3}{2}(1+w_{\phi}))}\frac{1}{(1+z_{r})^{3}}
\end{equation}
For $w_{\phi}=-0.96$ this becomes:
\begin{equation}
\left(\frac{M}{10^{-3}\mbox{ eV}}\right)^{4}>56\frac{1}{(1+z_{r})^{3}}
\end{equation}
In this case the allowed values of $z_{r}$ ($0< z_{r}<0.35$) can contribute only up to a factor of $1.5$, giving $M^{4}>5.6\times 10^{-11}\,\,\text{eV}^{4}$ for $z_{r}\rightarrow 0$ and $M^{4}>2.3\times 10^{-11}\,\,\text{eV}^{4}$ for $z_{r}=0.35$. As a consistency check, note that this last limit is compatible with the one derived for $z_{e}\le z_{r}$ in the limiting case $z_{e}=z_{r}=0.35$.
In Figure \ref{fig:M4ExclusionPlotsecondcase}  we also show a plot  with
 the allowed region for $M^{4}$ as a function of  the equation of state parameter $w_{\phi}$, considering the limiting cases of $z_{r}=0.35$ (blue region) and $z_{r}\rightarrow 0$ (purple region).

Hence we have found that the coupling scale $\Lambda$ must be much larger than the Planck scale. Expressed in terms of $f$ which is also a large scale determined by the acceleration of the Universe and which could appear as a natural cut-off for the model, $\Lambda$ must be around $10^6\sqrt{1+w_\phi}$ larger than $f$. Unless the equation of state is extremely close to $-1$ and the model becomes indistinguishable from a pure cosmological constant one, the direct coupling to photon must be extremely suppressed. This is a naturalness issue in these models which can only be resolved by embedding them in a larger framework. This is left for future work. On the other hand, the disformal scale $M$ is small and if identified with its value is some modified gravity models where $M=\sqrt{mM_{\rm Pl}}$ where $m$ is the graviton mass, we find that the graviton mass would have to be larger than the Hubble rate now, i.e. the graviton interaction would have a range of the order of the size of the Universe. We will investigate this possibility in future work.
\section{Conclusion}

Disformal couplings to matter naturally arise for scalar fields with shift symmetries, such as those suggested to explain the late time acceleration of the expansion of the Universe.  We have shown that such interactions can be constrained with cosmological observations, in particular with observations of the spectral distortions of the CMB and from tests of the distance duality relation. We have shown that both are affected by a disformal and a direct coupling to photons. The former leads to a variation of the speed of light and the latter a variation of the fine structure constant. We discussed how axionic quintessence models can lead to such variations and how they are constrained. In particular we have found that current bounds are compatible with a small variation of the speed of light, at the percent level,  since $z\sim 1$ at the onset of the acceleration of the Universe.
%
%

\acknowledgments
GG was supported in part by a
grant from the John Templeton Foundation. CB was supported by a Royal Society University Research Fellowship. ACD is supported in part by STFC.

\appendix

\section{Axionic Models}
\label{sec:app}
We focus on a typical example of quintessence model where the field eventually contributes to the acceleration of the Universe
For a large class of models, the field is frozen at the origin by Hubble friction until the Hubble rate decreases below the curvature of the potential. The subsequent evolution is a slow roll phase in order to have an equation of state $w\sim -1$ now. In particular, the field cannot move very much from its initial value as the kinetic energy could then start dominating its potential energy.
Take for instance as a typical example\footnote{This example is also discussed \emph{e.g} in \cite{Tsujikawa:2013fta}}
\be
V(\phi)=\frac{ \Lambda_0^4}{2}\left(1+ \cos\frac{\phi}{f}\right)\;,
\ee
where initially $\phi_i\ll f$. The initial tachyonic mass is $|m_i|^2= \frac{\Lambda_0^4}{f^2}$ and we require that the field starts rolling at $t_i$ where $m_i\sim H_r$ implying that we can choose
\be
\Lambda_0^4=2 {H_r^2 f^2}{}\;.
\ee
After $t_r$, the field starts rolling down. Linearising the
Klein Gordon equation around the origin, as we expect the field to move very little from its initial value,  we have
\be
\ddot \phi +3H \dot \phi -H_r^2 \phi=0\;.
\ee
Putting $\phi= a^{-3/2} u$ we have
\be
\ddot u -\left(\frac{3}{2} \dot H +\frac{9}{4} H^2 +H_r^2\right)u=0\;.
\ee
We find
\be
\ddot u-\left(H_r^2-\frac{9}{4} \Omega_\Lambda H^2 \omega_\phi\right) u=0\;,
\ee
where we have defined by $\Omega_\Lambda$ the dark energy fraction, $\omega_\phi$ the dark energy equation of state and we have not assumed that these quantities are constant.
Introduce the effective Hubble rate
\be
h^2(t)= H_r^2-\frac{9}{4} \Omega_\Lambda H^2 \omega_\phi\;,
\ee
which is a slowly varying function of time betwen $t_r$ and $t_0$.
As a result we can use the WKB approximation and get the solution
\be
u(t) =\frac{1}{\sqrt{h(t)}}\left(A e^{\int_{t_r}^t h(t') dt'}+Be^{-\int_{t_r}^t h(t') dt'}\right)\;,
\ee
where
\be
A-B= \frac{3}{2\sqrt{h(t_r)}}H_r u_i\;,
\ee
 and
\be
A+B= \sqrt{h(t_r)} u_i\;,
\ee
which is approximately (as we have linearised around the initial value)
\be
u(t)= u_i + \frac{3}{2} u_i H_r (t-t_r)\;,
\ee
and we have used $\dot \phi_r=0$ and therefore $\dot u_r= 3/2 H_r u_i$.
We have also
 \be
\dot u= \frac{3}{2} H_r u_i + H_r^2 u_i(1+\frac{9}{4}\Omega_{\Lambda r}) (t-t_r)\;.
\ee
The equation of state is close to -1 as long as the kinetic energy is much smaller than the potential energy, i.e. the field does not roll for long. If we identify the dark energy
\be
V(\phi_0)=\frac{\Lambda_0^4}{2}[1+\cos(\phi_0/f)]=  3\Omega_{\Lambda 0} H_0^2 m_{\rm Pl}^2\;,
\ee
this implies that
\be
\frac{\phi_0^2}{4f^2}= 1-\frac{3\Omega_{\Lambda 0} H_0^2 m_{\rm Pl}^2}{2 H_r^2 f^2}\;.
\ee
Notice that $\phi_0$ is small only provided that $f^2\approx \frac{3\Omega_{\Lambda 0} H_0^2 m_{\rm Pl}^2}{2H_{r}^2}$. In the following, we will parameterize the value
of the field now using the value of $f$.
We are now in a position to evaluate
\be
\dot \phi_0= \dot u_0 -\frac{3}{2} H_0 u_0\;,
\ee
as $a_0=1$.
This is also given by
\be
\dot\phi_0 =  H_r^2 u_i (t_0-t_r)\;.
\ee
We have then
\be
\dot\phi_0 = \frac{2}{3} (u_0-u_i) H_r\;,
\ee
or equivalently
\be
\dot \phi_0= \frac{2}{3} (\phi_0-a^{3/2}_r \phi_i) H_r\;.
\ee
Notice that the kinetic energy is given by
\be
T_\phi= \frac{1}{6} \Lambda_0^4 \frac{(\phi_0-a^{3/2}_r \phi_i)^2}{f^2}\;,
\ee
which is less than the potential energy $V(\phi_0)$ provided
\be
\frac{(\phi_0-a^{3/2}_r \phi_i)^2}{f^2}\ll 1\;,
\ee
as $\cos(\phi_0/f) \sim 1$.
This guarantees that the equation of state is close to $-1$. In fact we have
\be
w_\phi\approx -1+ 2\frac{T_\phi}{V(\phi_0)}\approx -1 +\frac{1}{6} \frac{(\phi_0-a^{3/2}_r \phi_i)^2}{f^2}\;.
\ee
As the initial condition is essentially at the origin $\phi_i=0$, we find that
\be
\phi_0= \sqrt{6} \sqrt{1+w_\phi} f \;,
\label{eq:Thawphi0}
\ee
and
the initial velocity is
\be
\dot \phi_0\approx  \frac{2}{3} \phi_0  H_0\;,
\ee
which determines the present value of $\phi_0$ as a function of the equation of state.
From this we can get that
\be
H_r^2 f^2=\frac{\frac{3}{2}\Omega_{\Lambda 0} H_0^2 m_{\rm Pl}^2}{1-\frac{3}{2}(1+w_\phi)}\;.
\ee
With these approximations, we find that the field evolves according to\be
\phi(t)=  \phi_0 \frac{t-t_r}{t_0-t_r}\;,
\ee
and
\be
\dot \phi= \frac{2}{3} \phi_0 H_0 \frac{t-t_r}{t_0-t_r}\;,
\ee
where in terms of the redshift
\be
z= H_0(t-t_0)\;,
\ee
and therefore
\be
\phi(z)=\phi_{0}\left(1-\frac{z}{z_{r}}\right)\;,
\label{eq:phiz}
\ee
and
\be
\dot \phi=\frac{2}{3}\phi_{0}H_{0}\left(1-\frac{z}{z_{r}}\right)\;.
\label{eq:dotphiz}
\ee
The field starts moving from the origin at a time $t_r$ determined by its redshift $z_r$ satisfying
\be
(1+z_r)^3=
\frac{\frac{3}{2}\Omega_{\Lambda 0}  m_{\rm Pl}^2}{f^2(1-\frac{3}{2}(1+w_\phi))}\;.
\label{eq:fsquareThawing}
\ee
The linear approximation is valid when $z_r \lesssim 1$. The model is determined by two parameters $w_\phi$ and $f$.
Hence, given an equation of state now $w_\phi$ and $f$ we get $\phi_0$ and $\dot\phi_0$.
This is enough to have
\be
\delta \phi= \phi_0 -\phi_i\approx \phi_0\;,
\ee
 and
\be
\delta\dot\phi= \dot\phi_0\;.
\ee
The fine structure constant is given by
\be
\frac{\alpha}{\alpha_\star}= 1 -\frac{4 \phi_0}{\Lambda} \left(1-\frac{z}{ z_r}\right)\;,
\label{eq:deltaAlphazr}
\ee
and the speed of light
\be
c_p^2 =1 -\frac{8 \phi_0^2 H_0^2}{9 M^4} \left(1-\frac{z}{ z_r}\right)^2\;,
\label{eq:deltaczr}
\ee
after $z_r$
This is equivalent to
\be
\frac{\alpha}{\alpha_\star}= 1 -4\sqrt{6} \sqrt{1+w_\phi}\frac{ f}{m_{\rm Pl}}\frac{m_{\rm Pl}}{\Lambda} \left(1-\frac{z}{ z_r}\right)\;,
\label{eq:deltaalphaf}
\ee
and the speed of light
\be
c_p^2 =1 -\frac{16(1+w_\phi)}{3 } \frac{H_0^2 m_{\rm Pl}^2}{M^4} \frac{f^2}{m_{\rm Pl}^2}\left(1-\frac{z}{ z_r}\right)^2\;,
 \label{eq:deltacf}
\ee
where $f/m_{\rm Pl}$ has been calculated previously.

¥
\end{document}